\def\mearth{{\rm\,M_\oplus}}
\def\msun{{\rm\,M_\odot}}

\documentclass{iaus}
\usepackage{graphicx}
\usepackage{deluxetable} 

\title[Terrestrial Planet Formation] 
{Terrestrial Planet Formation in Extra-Solar Planetary Systems}

\author[Sean N. Raymond]   
{Sean N. Raymond$^1$}

\affiliation{$^1$ NASA Postdoctoral Fellow, Center for Astrophysics and Space Astronomy and Center for Astrobiology, University of Colorado, Campus Box 389, Boulder, CO, 80309-0389, USA \\ email: {\tt raymond@lasp.colorado.edu} }

\pubyear{2008}
\volume{249}  
\jname{Exoplanets: Detection, Formation and Dynamics}
\editors{eds.}
\begin{document}

\maketitle

\begin{abstract}
Terrestrial planets form in a series of dynamical steps from the solid component of circumstellar disks. First, km-sized planetesimals form likely via a combination of sticky collisions, turbulent concentration of solids, and gravitational collapse from micron-sized dust grains in the thin disk midplane. Second, planetesimals coalesce to form Moon- to Mars-sized protoplanets, also called ``planetary embryos''. Finally, full-sized terrestrial planets accrete from protoplanets and planetesimals. This final stage of accretion lasts about 10-100 Myr and is strongly affected by gravitational perturbations from any gas giant planets, which are constrained to form more quickly, during the 1-10 Myr lifetime of the gaseous component of the disk. It is during this final stage that the bulk compositions and volatile (e.g., water) contents of terrestrial planets are set, depending on their feeding zones and the amount of radial mixing that occurs.  The main factors that influence terrestrial planet formation are the mass and surface density profile of the disk, and the perturbations from giant planets and binary companions if they exist.  Simple accretion models predicts that low-mass stars should form small, dry planets in their habitable zones.  The migration of a giant planet through a disk of rocky bodies does not completely impede terrestrial planet growth.  Rather, "hot Jupiter" systems are likely to also contain exterior, very water-rich Earth-like planets, and also "hot Earths", very close-in rocky planets.  Roughly one third of the known systems of extra-solar (giant) planets could allow a terrestrial planet to form in the habitable zone. 
\end{abstract}

\firstsection 
\section{Introduction}

Recent research has developed a model for the growth of the terrestrial planets in the Solar System via collisional accumulation of smaller bodies (e.g., Wetherill 1990).  Several distinct dynamical stages have been identified in this process, from the accumulation of micron-sized dust grains to the late stage of giant impacts.  In this article, I first review the current state of knowledge of the stages of terrestrial planet formation ($\S$ 2).  Second, I explore the effects of external parameters on the accretion process, including new work showing that simple accretion models predict that low-mass stars should preferentially harbor low-mass terrestrial planets in their habitable zones ($\S$ 3).  These include the orbit of a giant planet and the surface density profile of the protoplanetary disk.  Third, I examine the effects of the migration of a Jupiter-mass giant planet on the accretion of terrestrial planets ($\S$ 4), and also apply accretion models to the known set of extra-solar planets to show that roughly one third of the known systems could have formed an Earth-like planet in the habitable zone ($\S$ 5).  Finally, conclusions and avenues for future study are presented ($\S$ 6).  

\section{Stages of Terrestrial Planet Growth}

The process of planet formation starts when a parcel of gas with a giant molecular cloud becomes gravitationally unstable, and completes on a $\sim 10^8$ year timescale with the formation of a complete planetary system.  I cannot hope to encompass all relevant processes in this short article.  Therefore, I will start from a protoplanetary disk in which dust has settled to the thin "mid-plane" of the disk, a process which requires $\sim10^4$ years in a non-turbulent disk (Weidenschilling 1980).  I will review three distinct dynamical stages of terrestrial planet growth: (i) the formation of km-sized planetesimals starting from dust grains; (ii) the formation of $\sim$1000-km sized protoplanets (also called "planetary embryos"); and (iii) the formation of full-sized terrestrial planets.   For more detailed reviews, the reader is referred to Lissauer (1993), Chambers (2004), and Papaloizou \& Terquem (2006).  

Models suggest that the surface density of solid material $\Sigma$ in disks roughly follow a power law with radial distance $r$, i.e.,  
\begin{equation}
\Sigma = \Sigma_1 r^{-\alpha},
\end{equation}
\noindent where $\Sigma_1$ is the surface density at 1 AU and $\alpha$ controls the radial distribution of solids (not to be confused with the viscosity parameter of the same name).  In the "minimum-mass solar nebula" (MMSN) model of Weidenschilling (1977a) and Hayashi (1981), $\alpha$ = 3/2.  However, new analyses of the MMSN model derive values for $\alpha$ between 1/2 and 2 (Davis 2005; Desch 2007).  In addition, current models and observations generally favor $\alpha \approx 1$ (Dullemond \etal 2007; Andrews \& Williams 2007; Garaud \& Lin 2007).  

{\bf From dust to planetesimals.}  The growth of km-sized planetesimals from micron-sized dust grains presents a significant modeling challenge, which can be constrained to some degree by observations of dust populations in disks around young stars (Dullemond \& Dominik 2005).  Collisional growth of micron-sized grains, especially if they are arranged into fluffy aggregates, appears efficient for relatively small particle sizes and impact speeds of $\sim 1 \, m\, s^{-1}$ or slower (Dominik \& Tielens 1997; Wurm \& Blum 2000; Poppe \etal 2000; Benz 2000; see Dominik \etal 2007 and references therein).  However, there is a constant battle between disk turbulence, which increases random velocities, and drag-induced settling, which reduces them (Cuzzi \etal 1993).  Growth of particles in such collisions appears effective until they reach roughly 1 m in size.  At that point, continued growth may be suppressed by collision velocities of $\gtrsim 10 \, m\, s^{-1}$ (Dominik \etal 2007), as well as rapid inspiralling of m-sized bodies due to aerodynamic drag (Weidenschilling 1977b).  Indeed, the timescale for infall can be as short as 100 years, leading to what has been called the "meter-size catastrophe".  Collisional growth models must quickly cross the barrier at meter-sizes (Weidenschilling \& Cuzzi 1993; Benz 2000; Weidenschilling 2000).  

If solids can be sufficiently concentrated in the disk relative to the gas, then local gravitational instability may occur, leading to the top-down formation of planetesimals (Goldreich \& Ward 1973; Youdin \& Shu 2002).  Concentration of particles has been proposed to occur at local maxima in the disk pressure (Haghighipour \& Boss 2003), which can exist as a result of turbulence in the disk (Johansen \etal 2007).  It is interesting to note that m-sized particles are the fastest to drift toward pressure maxima (Johansen \etal 2006), for the same reason as the "meter-size catastrophe."  An alternate location for planetesimal formation via gravitational instability are regions with an increased local density of solids (Goodman \& Pindor 2000), via concentration due to drag-induced inspiralling (Youdin \& Chiang 2003), in vortices (Tanga \etal 1996), or even via photo-evaporative depletion of the gas layer (Throop \& Bally 2005).  

Thus, the details of planetesimal formation are not fully resolved.  The current best understanding invokes a combination of collisional sticking of grains at small sizes until they reach $\sim$ 1 m in size, concentration of these boulders in pressure maxima, and subsequent growth, via either gravitational instability or collisional accretion, into km-sized planetesimals.  These objects are the building blocks of terrestrial planets.

{\bf From planetesimals to protoplanets.}  While gas is still present in the disk, eccentricities of planetesimals are damped quickly due to aerodynamic gas drag (Adachi \etal 1976).  While velocities remain low, bodies that are slightly larger than the typical size can accelerate their growth due to gravitational focusing (Safronov 1969; Greenberg \etal 1978):
\begin{equation}
\frac{dM}{dt} = \pi R^2 \left(1+\frac{v_{esc}^2}{v_{rand}^2}\right),
\end{equation}
\noindent where $R$ represents the body's physical radius, $v_{esc}$ is the escape speed from the body's surface (= $\sqrt{2 G M/R}$), and $v_{rand}$ represents the velocity dispersion of planetesimals.  While random velocities are small, gravitational focusing can increase the growth rates of bodies by a factor of hundreds, such that $dM/dt \sim M^{4/3}$, leading to a phase of "runaway growth" (Greenberg \etal 1978; Wetherill \& Stewart 1989, 1993; Ida \& Makino 1992; Kokubo \& Ida 1996; Goldreich \etal 2004).  The length of this phase depends on the timescale for random velocities of planetesimals to approach the escape speed of the larger bodies.  For small ($\sim$100 m-sized) planetesimals, gas drag is stronger such that runaway growth can be prolonged and embryos may be larger and grow faster (Chambers 2006).  

Eventually, the random velocities of planetesimals are increased by gravitational interactions with the larger bodies that have formed, in a process called "viscous stirring" (Ida \& Makino 1992a).  During this time, the random velocities of large bodies are kept small via dynamical friction with the swarm of small bodies (Ida \& Makino 1992b).  Gravitational focusing is therefore reduced, and the growth of large bodies is slowed to the geometrical accretion limit, such that $dM/dt \sim M^{2/3}$ (Ida \& Makino 1993; Rafikov 2003).  Nonetheless, large bodies continue to grow, and jostle each other such that a characteristic spacing of several mutual Hill radii $R_{H, m}$ is maintained ($R_{H,m} = 0.5 [a_1+a_2]\,[M_1+M_2/3M_\star]^{1/3}$, where $a_1$ and $M_1$ denote the orbital distance and mass of object 1, etc; Kokubo \& Ida 1995).  This phase of growth is often referred to as "oligarchic growth", as just a few large bodies dominate the dynamics of the system, with reduced growth rates and increased interactions between neighboring protoplanets (Kokubo \& Ida 1998; Leinhardt \& Richardson 2005).  

\begin{figure}
\begin{center}
 \includegraphics[width=5in]{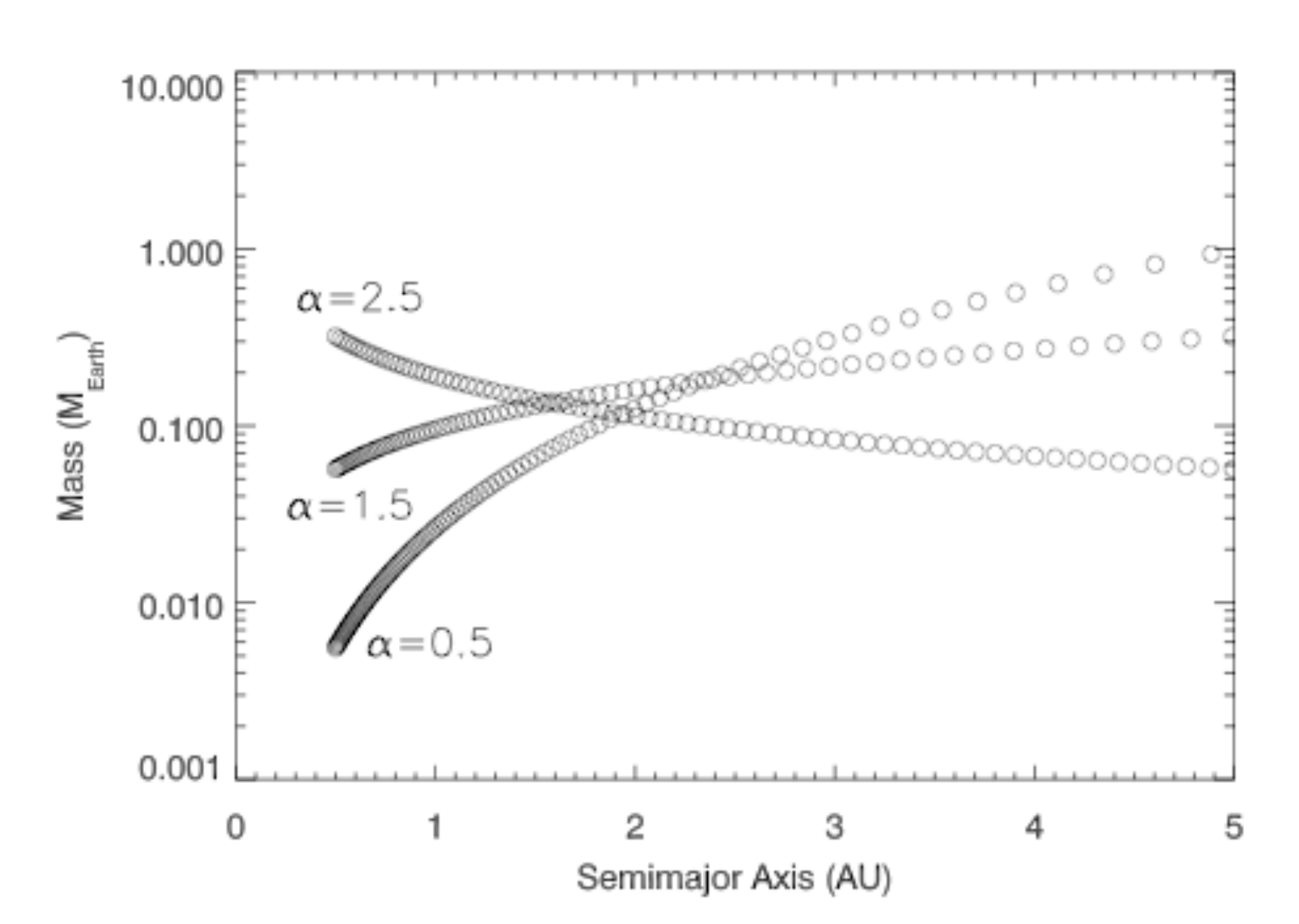} 
 \caption{Three sets of protoplanets at the end of oligarchic growth, assuming they formed with separations of 7 mutual Hill radii.  Each set of protoplanets contains 10 $\mearth$ between 0.5 and 5 AU, differing only in the radial surface density exponent $\alpha$.  From Raymond \etal (2005b).}
   \label{fig:alpha}
\end{center}
\end{figure}

Oligarchic growth tends to form systems of protoplanets with roughly comparable masses and separations of 5-10 mutual Hill radii (Kokubo \& Ida 1998, 2000; Weidenschilling \etal 1997).  Masses and separations of protoplanets depend on the total mass and surface density distribution of the disk (Kokubo \& Ida 2002).  Typical protoplanet masses in a solar nebula model are a few percent of an Earth mass, i.e., roughly lunar to Mars-sized (Kokubo \& Ida 2000; Collis \& Sari 2007).  Fig~\ref{fig:alpha} shows three distributions of protoplanets: each contains 10 $\mearth$ between 0.5 and 5 AU (roughly twice the minimum-mass disk), with 7 $R_{H,m}$ between adjacent protoplanets (from Raymond \etal 2005b).  Note that for surface density profiles steeper than $r^{-2}$, the protoplanet mass decreases with orbital distance.

It is important to realize that terrestrial protoplanets are the same objects as giant planet cores, assuming giant planets to form via the bottom-up, "core-accretion" scenario (Mizuno 1980; Pollack \etal 1996; Ida \& Lin 2004; Alibert \etal 2005).  Models of core-accretion estimate the growth rate of an isolated core, then calculate the accretion of gas onto the core, generally neglecting the potentially important effects of nearby cores.  Note also that protoplanets may excite spiral density waves in the gaseous disk and undergo inward type 1 migration on a 10$^{4-5}$ year timescale (Goldreich \& Tremaine 1980; Ward 1986; Masset \etal 2006).  This can reduce the efficiency of terrestrial planet growth by removing a significant fraction of material (McNeil \etal 2005; Ida \& Lin 2007).  

{\bf From protoplanets to terrestrial planets.}  Once the local mass in planetesimals and protoplanets (i.e., "small" and "large" bodies, respectively) is comparable, feeding zones of neighboring protoplanets overlap, and giant collisions between protoplanets begin to occur (Wetherill 1985; Kenyon \& Bromley 2006).  This is the start of the final phase of terrestrial planet growth, sometimes called "late-stage accretion" (Wetherill 1996; Chambers \& Wetherill 1998).  During this stage, planets grow by accreting other protoplanets as well as planetesimals (Chambers 2001; Raymond \etal 2006a; O'Brien \etal 2006).  Eccentricities of protoplanets are generally smaller than for planetesimals because of dynamical friction, but not small enough to prevent protoplanet collisions.  Growing planets clear their nearby zones of material, and their feeding zones widen and move outward in time (Raymond \etal 2006a).  During the very late stages of accretion, there can be substantial mixing across the inner planetary system, with important implications for terrestrial planet compositions (see below). 

Giant collisions between protoplanets may not always be accretionary: high-speed or off-center collisions can actually erode the target mass (Agnor \& Asphaug 2004; Asphaug \etal 2006; Genda \etal 2007, in preparation).  High-speed collisions may alter the planetary composition, by preferentially removing more volatile materials; this is a proposed explanation for Mercury's large iron core (Wetherill 1988; Benz \etal 1988).  In addition, low-speed, off-center collisions have the potential to create a circum-planetary disk of debris from which a large moon may accrete (Benz \etal 1987; Ida \etal 1997; Canup \& Asphaug 2001; Canup 2004).  Note, however, that such low-speed collisions are rare, especially at later times when a planet's nearby zone has been cleared out (Agnor \etal 1999).  

Note that giant planets are constrained to form in the 1-10 Myr lifetimes of the gaseous component of protoplanetary disks (Haisch \etal 2001; Brice\~no \etal 2001; Pascucci \etal 2006).  Thus, if giant planets form in a planetary system, as Jupiter and Saturn did in the Solar System, then they can have an effect on the late-stage accretion of terrestrial planets.  Indeed, it has been shown that giant planets have a large impact on the final assembly of terrestrial planets (Levison \& Agnor 2003; Raymond \etal 2004; Raymond 2006; O'Brien \etal 2006).  For example, in systems with less massive giant planets, induced eccentricities are smaller, reducing the width of feeding zones and causing the formation of more, smaller terrestrial planets as compared with a more massive giant planet.  In addition, in systems with more massive or eccentric giant planets, the innermost terrestrial planet tends to be the most massive (Levison \& Agnor 2003).  

\begin{figure}
\begin{center}
 \includegraphics[width=5in]{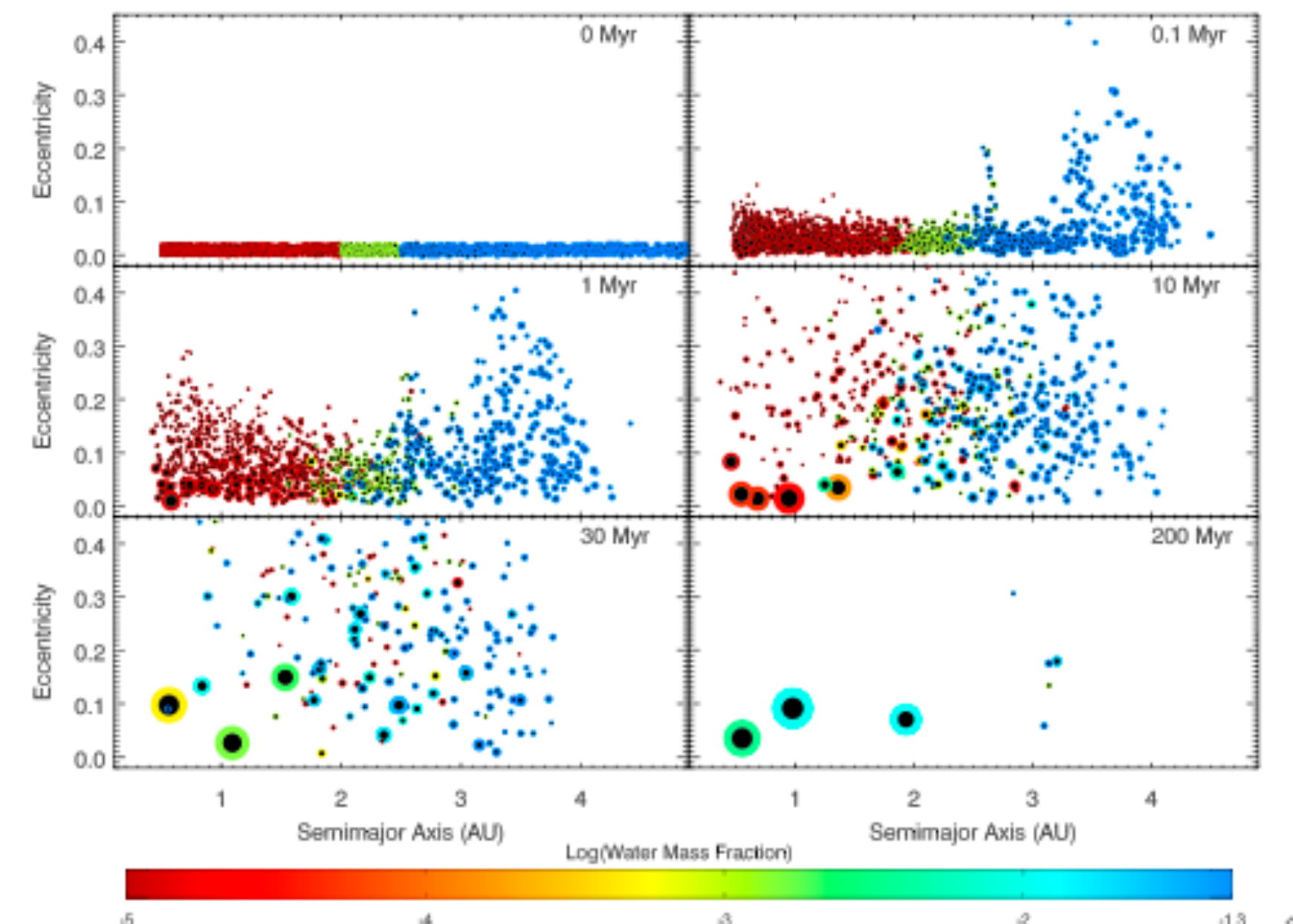} 
 \caption{Snapshots in time from a simulation of the final growth of terrestrial planets, starting from 1885 sub-isolation mass objects (from Raymond \etal 2006a).  The size of each body is proportional to its mass$^{1/3}$, the dark circle represents the relative size of each body's iron core (in the black and white version, iron cores are shown only for bodies larger than 0.05 $\mearth$), and the color corresponds to its water content (red = dry, blue = 5\% water; in the black and white version, white = dry and black = 5\% water; see color bar).  For a movie of this simulation, go to http://lasp.colorado.edu/$\sim$raymond/ and click on ``movies and graphics''.}
   \label{fig:insitu}
\end{center}
\end{figure}

Figure~\ref{fig:insitu} shows snapshots in time of the accretion of terrestrial planets from a simulation by Raymond \etal (2006a), including 9.9 $\mearth$ of material from 0.5-5 AU (roughly twice the mass of the MMSN model).  The simulation included a Jupiter-mass giant planet on a circular orbit at 5.5 AU, which is somewhat different than its current orbit, although it is consistent with the 'Nice' model of the giant planets' orbital evolution (Tsiganis \etal 2005). Several mean motion resonances are clearly visible as vertical spikes in eccentricity in the '0.1 Myr' panel.  Eccentricities are driven in the inner disk via interactions between protoplanets, and in the outer disk via secular and resonant perturbations from the giant planet.  Large bodies grow more quickly closer to the Sun, because of the faster orbital timescales; by 10 Myr the planet at $\sim$ 1 AU has reached 1 $\mearth$.  However, large-scale mixing between zones does not happen until about 20 Myr, when the feeding zones of all three final planets overlap and encompass the entire terrestrial zone, out to $\sim$4 AU (Raymond \etal 2007a).  The final configuration of three planets contains a 1.5 $\mearth$ planet at 0.55 AU, a 2 $\mearth$ planet at 0.98 AU, and a 0.95 $\mearth$ planet at 1.93 AU (see Raymond \etal 2006a for details).  The orbits of the planets have slightly higher eccentricities than the current-day terrestrial planets; in similar simulations, O'Brien \etal (2006) formed terrestrial planet systems with eccentricities even lower than those of Earth, Venus and Mars.  

Formation times of Earth-like planets from dynamical simulations are 10-100 Myr, comparable to timescales derived from Hf/W and other isotopic systems for the last core-forming event on Earth (e.g., Kleine \etal 2002; Jacobsen 2005).  Although this formation timescale is far longer than the gaseous disk lifetime, Interactions with the dissipating gaseous disk may affect this final stage of accretion.  Important processes include tidal gaseous drag (also called "type 1" damping; Ward 1993; Agnor \& Ward 2002; Kominami \& Ida 2002, 2004) and secular resonance sweeping, which can be induced by the changing gravitational potential as the disk dissipates (Ward 1981; Nagasawa \etal 2005).  

The final compositions of terrestrial planets are determined mainly during this final stage of terrestrial accretion.  A planet's composition is mainly determined by its feeding zone, i.e., the sum of all the material incorporated during formation (as well as physical processes during and after accretion).  If planets form locally, then their compositions are a simple reflection of the composition of the local building blocks.  Thus, eccentricity-driven mixing between different radial zones during accretion is perhaps the key process that determines the final planetary composition.

The Earth's water content is anomalously high: nebular models suggest that the local temperature at 1 AU was too hot to allow for hydration of planetesimals (Boss 1998).  Thus, it is thought that Earth's water was "delivered" from more distant regions, in the form of hydrated asteroids (Morbidelli \etal 2000; Raymond \etal 2004, 2007b) or comets (Owen \& Bar-Nun 1995).  The D/H ratio of Earth's water is virtually identical to that of carbonaceous chondrites, which are linked to C type asteroids in the outer main belt (Robert \& Epstein 1982; Kerridge 1985; see Table 1 of Morbidelli \etal 2000).  Comets, though poorly sampled, appear to have D/H ratios two times higher than Earth (Balsiger \etal 1995; Meier \etal 1998; Bockelee-Morvan \etal 1998), suggesting that primitive outer asteroid material may be the best candidate for the source of Earth's water.  In Fig.~\ref{fig:insitu}, material is given a starting water content that matches the values for primitive meteorites (Abe \etal 2000; see Fig.~4 of Raymond \etal 2004).  Radial mixing during formation delivers water from the primitive outer asteroid belt (beyond 2.5 AU) to the growing terrestrial planets.  The amount of water delivered in planetesimals vs. protoplanets is comparable (Raymond \etal 2007a). Thus, we expect that water delivery from planetesimals is statistically robust, while the amount of additional water from a small number of water-rich protoplanets can vary significantly from system to system.  

\section{Effects of External Parameters on Terrestrial Accretion}

Dynamical simulations of terrestrial planet growth display a wide range in outcomes, in terms of planet masses, orbits and compositions (Wetherill 1996; Agnor \etal 1999; Chambers 2001; Raymond \etal 2004).  The stochastic nature of accretion is due to the importance of individual scattering events during the very late phases of accretion.  Indeed, Quintana \& Lissauer (2006) showed that varying the initial conditions by moving a single embryo by just 1 meter in its same orbit could cause a significant change in the outcome.  Nonetheless, systematic trends are seen between certain system parameters and the properties of the planets that form.  Many simulations with a given starting condition are needed to suppress stochastic effects and to understand the systematic trends.

To date, systematic trends have been established with the following system parameters: 1) the total disk mass; 2) the disk's surface density profile; 3) the stellar mass; 4) the mass and 5) orbit of external giant planets; and 6) the mass and 7) separation of a binary stellar companion.  Here I summarize the effect of these parameters on terrestrial planet formation, in terms of the terrestrial planets' final masses, orbits, and compositions (specifically water contents, in the context of water delivery from an asteroidal source, as described in $\S$ 2).  

{\bf Disk Mass.}  There exists a range of roughly two orders of magnitude in observed circumstellar disk masses (Andre \& Montmerle 1994; Eisner \& Carpenter 2003; Andrews \& Williams 2005; Scholz \etal 2006).  An increase in the disk mass allows more massive embryos to form than a fiducial case (Kokubo \& Ida 2002).  These embryos have stronger gravitational interactions, leading to larger eccentricities and therefore wider feeding zones.  Thus, a disk with a larger total mass in solids will tend to form a smaller number of more massive terrestrial planets than a less massive disk (Raymond \etal 2005a, 2007; Kokubo \etal 2006).  The planet mass, $M_p$, scales with the disk mass, $M_d$, as $M_d^{1.1}$ (Kokubo \etal 2006; Raymond \etal 2007b).  The super-linear scaling is due to the increase in feeding zone width.  In addition, a wider feeding zone implies increased radial mixing, such that more massive disks tend to form more water-rich planets in the habitable zone (Raymond \etal 2007b).

{\bf Disk Surface Density Profile.}  The surface density profile of protoplanetary disks is uncertain; estimates range from $\Sigma \sim r^{-1/2}$ to $r^{-2}$ (i.e., from eqn. 2.1, $\alpha$ = 0.5-2; e.g., Weidenschilling 1977a; Davis 2005; Kuchner 2004; Desch 2005; Andrews \& Williams 2007).  Systematic simulations of disks with varying surface density profiles were run by Raymond \etal (2005b) and Kokubo \etal (2006).  They found that for steeper profiles (higher values of $\alpha$), the terrestrial planets (1) are more numerous, (2) form more quickly, (3) form closer to the star, (4) are more massive, (5) have higher iron contents, and (6) have lower water contents. However, the possibility of forming potentially habitable planets (water-rich planets in the habitable zone) does not vary strongly with $\alpha$.

{\bf Stellar Mass.}  A very rough correlation appears to exist between protoplanetary disk mass and the mass of the host star, with a scatter of about 2 orders of magnitude for a given stellar mass (e.g., Fig.~3 of Scholz \etal 2006).  Indeed, eqn. 2.1 can be rewritten as:
\begin{equation}
\Sigma = \Sigma_1 \, f \, Z \, r^{-\alpha} \left(\frac{M_\star}{M_\odot}\right)^h,
\end{equation}
where $f$ represents the relative disk mass, $Z$ represents the disk metallicity on a linear scale (i.e., how much of the disk mass is in solids), and $M_\star$ is the stellar mass (Ida \& Lin 2005; Raymond \etal 2007b).  Current best-fit values are $\alpha \sim 1$ (as discussed above) and $h \sim 1$, while $f$ and $Z$ are unique to each disk and have inherent distributions (e.g., Andrews \& Williams 2005; Nordstrom \etal 2004).  Given that terrestrial planets are thought to form interior to the snow line, the location of terrestrial planet formation depends on the stellar luminosity, which is a strong function of the stellar mass (e.g., Hillenbrand \& White 2004).  Indeed, the habitable zones of low-mass stars are very close-in (Kasting \etal 1993), where the amount of material available to form terrestrial planets is limited.

\begin{figure}
\begin{center}
 \includegraphics[width=5in]{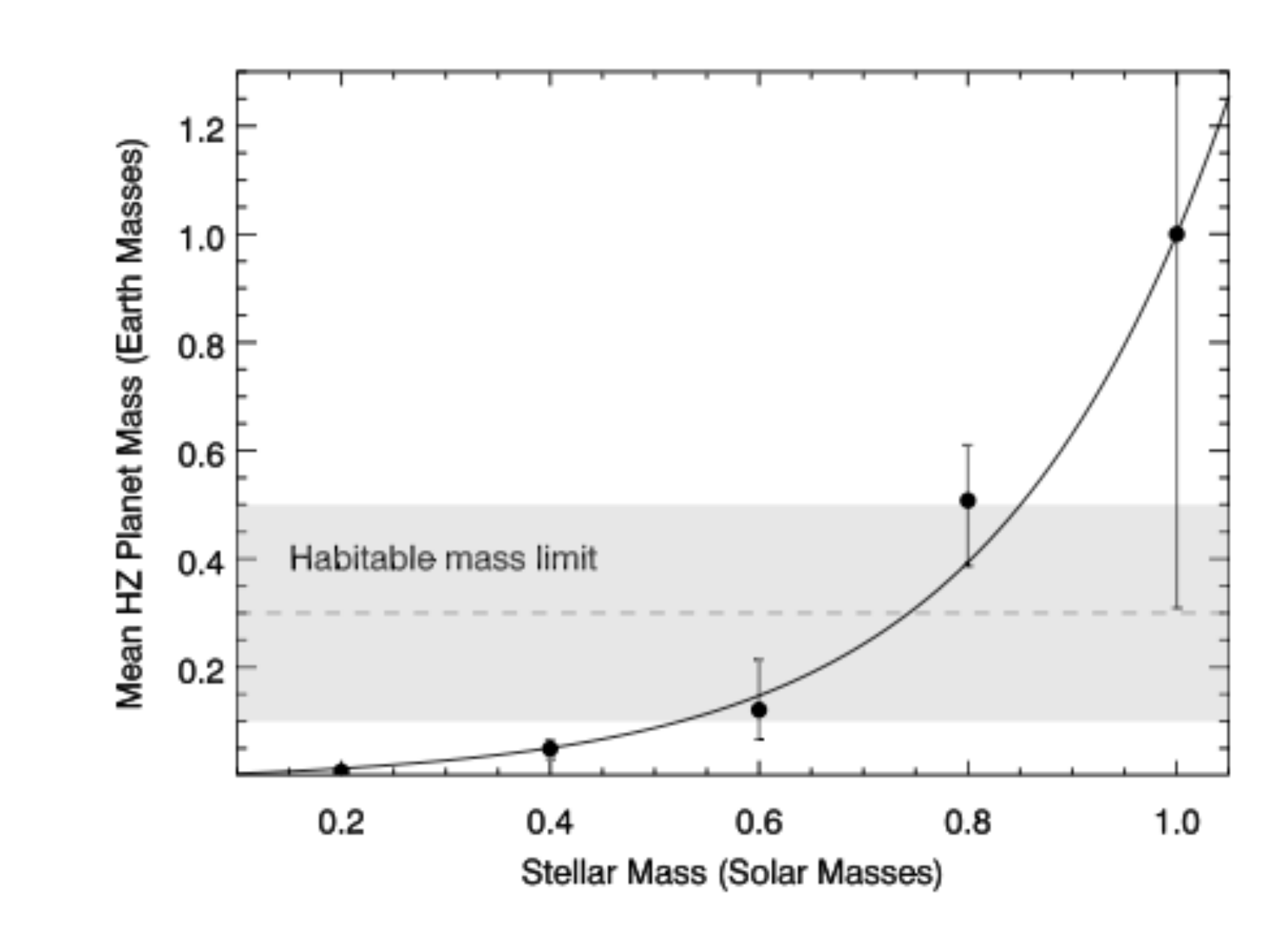} 
 \caption{An estimate of the mass of terrestrial planets formed via accretion in the habitable zone as a function of stellar mass, assuming $f=\alpha=h=1$ (see eqn.3.1).  From Raymond \etal (2007b).  }
   \label{fig:mstars}
\end{center}
\end{figure}

Figure~\ref{fig:mstars} shows the expected mean mass of terrestrial planets in the habitable zone as a function of stellar mass in the fiducial model of Raymond \etal (2007b), with $f Z$ = 1, $h$ = 1, and $\alpha$ = 1.  The data come from 20 simulations of the late-stage accretion of terrestrial planets from protoplanets, with the error bars signifying the range of outcomes.  The curve represents a very simple model, in which the mass of a terrestrial planet scales linearly with the local mass (as expected from results of Kokubo \etal 2006).  Both the curve and data are calibrated to match 1 Earth mass for a Solar-mass star.  The "habitable planet limit" of 0.1-0.5 $\mearth$ represents the range of expected planet masses capable of sustaining life for long timescales (Williams \etal 1997; Raymond \etal 2007b).  

Raymond \etal (2007b) therefore show that a simple {\it in situ} accretion model predicts that low-mass stars should typically harbor low-mass terrestrial planets in their habitable zones.  This result is contingent on the parameters of eqn. 3.1, especially the value of $h$.  In addition, this study neglects interactions with the gaseous disk as well as tidal effects, both of which could be important for terrestrial planets around low-mass stars (Ida \& Lin 2005; Barnes \etal 2007).  If many large terrestrial planets are discovered in the habitable zones of low-mass stars, the study of Raymond \etal (2007b) provides evidence that these planets did not form via {\it in situ} accretion, but rather migrated to their current locations (see, e.g., Terquem \& Papaloizou 2007).  In fact, it may be possible to differentiate planet formation models in systems with close-in terrestrial planets if the planet transits its host star, given the relation between planet size and composition (Valencia \etal 2007; Seager \etal 2007; Fortney \etal 2007; Selsis \etal 2007a; Raymond \etal 2007c).  

{\bf Giant Planet Mass.}  The effects of giant planet mass mimic those of the disk mass in some ways.  A more massive giant planet excites larger eccentricities among protoplanets, leading to wider feeding zones and therefore fewer, more massive terrestrial planets (Levison \& Agnor 2003; Raymond \etal 2004).  In addition, large eccentricities among embryos often causes the innermost terrestrial planet to be the most massive (Levison \& Agnor 2003).  However, giant planets in virtually all configurations are destructive to asteroidal water delivery (Raymond 2007, in preparation).  The larger eccentricities induced by a more massive giant planet cause a larger fraction of water-rich material (2-4 AU) to become unstable and be removed from the system, usually via ejection (Raymond 2007, in preparation).  

{\bf Giant Planet Orbit.}  The current explanation for the deficit of mass in the asteroid belt is that the material was removed because of the dynamical effects of Jupiter (Wetherill 1992; Petit \etal 2001; Bottke \etal 2005).  There exists a minimum separation between a giant planet and the closest terrestrial planet that can form, of a factor of 3-5 in orbital period (Raymond \etal 2005b; Raymond 2006).  Thus, terrestrial planet formation is inhibited in the vicinity of a giant planet, such that a giant planet's orbital distance provides information as to where terrestrial planets could have formed in the system (see $\S$ 5 and Raymond 2006).  In addition, for giant planets on more distant orbits, more water-rich material is available to the terrestrial planet region, allowing planets to be somewhat more water-rich, although this effect is small compared with the disk mass (Lunine 2001; Raymond \etal 2004, 2007b). 

The orbital eccentricity of a Jupiter-like planet has important consequences for water delivery to terrestrial planets.  An eccentric giant planet destabilizes primitive, water-rich asteroidal material and prevents it from colliding with growing planets in the inner system (Chambers \& Cassen 2002; Raymond \etal 2004; Raymond 2006; O'Brien \etal 2006).  Thus, dry terrestrial planets are probably correlated with eccentric giant planets.

{\bf Binary Companion Mass and Separation.}  Terrestrial planet formation has been studied in binary star systems, both orbiting both components of a close binary (P-type orbits) or one star with a more distant companion (S-type orbits).  Perturbations from the binary can increase the relative velocities of planetesimals, making it difficult to form protoplanets (Thebault \etal 2002).  However, gas drag can act to align the orbital apses of planetesimals, thereby reducing collision velocities and allowing protoplanets to grow (Thebault \etal 2006).  This effect is also seen for protoplanet formation in the presence of a giant planet (Kortenkamp \etal 2001).  

Terrestrial planets on P-type orbits are statistically the same if the binary's maximum separation is less than 0.2 AU (Quintana \& Lissauer 2006).  For larger separations, fewer terrestrial planets form, especially in the inner disk.  Water delivery in these systems has not been studied.  However, for an equal mass binary of two 0.5 $\msun$ stars, the habitable zone is actually much closer in than for the Sun.  If the habitable zone scales with the stellar flux, i.e., as the square root of the stellar luminosity, then the total luminosity of two 0.5 $\msun$ stars is only 6\% of the Solar flux, using the mass-luminosity relation of Scalo \etal (2007), which is a fit to data of Hillenbrand \& White (2004).  Thus, if the Sun's habitable zone is at 0.8-1.5 AU, then the habitable zone of this binary system is at 0.2-0.37 AU, such that it is far more vulnerable to the effects of the binary than a na\"ive contemplation.

As for the case of a giant planet's orbital distance, there exists a correlation between the orbit of a binary companion for S-type orbits and the most distant terrestrial planet that can form.  For an equal-mass binary, there exists a ratio of roughly 5 between the orbital periastron of the binary and the more distant terrestrial planet that can form around the primary (Quintana \etal 2007; Haghighipour \& Raymond 2007).  Indeed, Quintana \etal (2002) showed that terrestrial planets could accrete in the $\alpha$ Centauri binary system.  Thus, binaries with separations larger than 5-10 AU do not directly impede the accretion of terrestrial planets, although they may act to preferentially remove water-rich material as it lies at larger orbital distances and is therefore more susceptible to binary perturbations.  In cases with a giant planet on a Jupiter-like orbit and a binary companion, the giant planet can act to transfer angular momentum between the binary star and inner terrestrial material (Haghighipour \& Raymond 2007).  The binary's orbit can induce a large eccentricity in the giant planet and cause it to deplete water-rich material as discussed in the case of an isolated, eccentric giant planet.

\section{Giant Planet Migration and Terrestrial Planet Formation}

Close-in giant planets are thought to have formed at larger orbital distances and migrated in to their current locations (Lin \etal 1996; Bodenmeimer \etal 2000).  Planets more massive than roughly a Saturn-mass have feeding zones that are wider than the disk height, and so carve annular gaps in the disk (Lin \& Papaloizou 1986; Takeuchi \etal 1996; Rafikov 2002; Crida \etal 2006).  Such planets are then tied to the viscous evolution of the disk, and ``type 2'' migrate, usually inward, on the viscous timescale of $\sim 10^{5-6}$ years (Lin \& Papaloizou 1986; Ward 1997; D'Angelo \etal 2003; Ida \& Lin 2004).  On its way to becoming a hot Jupiter or falling into the star, the giant planet must pass through the terrestrial planet zone.  

Until recently, the effects of this migration on the accretion of Earth-like was debated without any direct calculations.  Several authors made unfounded, pessimistic assumptions about the existence of terrestrial planets (Ward \& Brownlee 2000; Gonzalez \etal 2001; Lineweaver \etal 2004).  Others made limited calculations with mixed results, both pessimistic (Armitage 2003; Mandell \& Sigurdsson 2003) and optimistic (Edgar \& Artymowicz 2004; Raymond \etal 2005a; Lufkin \etal 2006).  Since 2005, two groups have conducted realistic simulations of the migration of a giant planet through a region of accreting terrestrial bodies, with similar results.  These papers include the migration of a single giant planet, a disk containing more than a thousand terrestrial particles, and effects of the gaseous disk on the terrestrial bodies (Fogg \& Nelson 2005, 2007a, 2007b; Raymond \etal 2006b; Mandell \etal 2007).  

\begin{figure}
\begin{center}
 \includegraphics[width=5in]{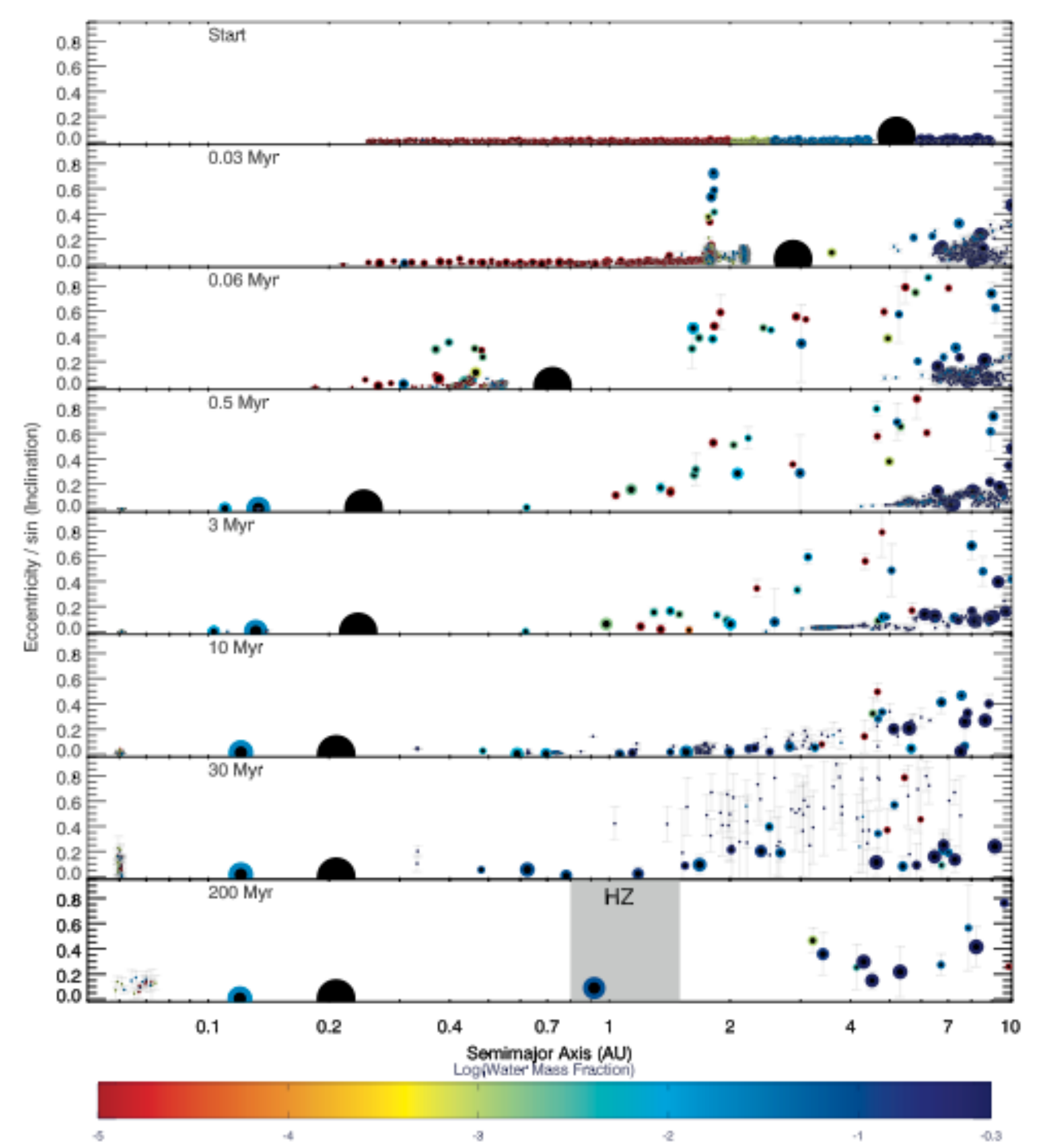} 
 \caption{Snapshots in time from a simulation of the growth of terrestrial planets in the presence of giant planet migration.  The giant planet is shown in black, and note the logarithmic scale of the x axis.  As in Fig.~\ref{fig:insitu}, the size of each body is proportional to its mass$^{1/3}$, the dark circle represents the relative size of each body's iron core (color version only), and the color corresponds to its water content (red = dry, dark blue = 50\% water; in the black and white version white = dry and black = 50\% water; see color bar).  For bodies with inclinations larger than 5$^{\rm o}$, inclinations are shown via vertical error bars with lengths of sin (inc.), readable on the y axis.  Reproduced from Raymond \etal (2006b).  For a movie of this simulation, go to http://lasp.colorado.edu/$\sim$raymond/ and click on ``movies and graphics''.}
   \label{fig:mig}
\end{center}
\end{figure}

Figure~\ref{fig:mig} shows snapshots of a simulation that was integrated for 200 Myr, including the migration of a Jupiter-mass giant planet from 5.2 to 0.25 AU in the first 100,000 years, from Raymond \etal (2006b).  During the giant planet's migration, material interior to the giant planet is shepherded inward by mean motion resonances (MMRs), especially the 2:1 MMR (Tanaka \& Ida 1999).  This shepherding occurs because eccentricities are increased by the MMR and subsequently damped by gas drag and dynamical friction, causing a net loss of orbital energy.  In this simulation, roughly half of the total material was shepherded inward, while the other half underwent a close encounter with the giant planet and was thrown outward, exterior to the giant planet.  This material had very high eccentricities and inclinations, but these were damped on Myr timescales by gas drag (Adachi \etal 1976) and dynamical friction.  On 100 Myr timescales, accretion continued in the outer disk, including the formation of a 3 $\mearth$ planet at 0.9 AU.  This planet orbits its star in the habitable zone, and has roughly 30 times more water than other simulations designed to reproduce the Earth (Raymond \etal 2004).  Thus, this planet is likely a ``water world'' or ``ocean planet'', with deep oceans covering its surface (L\'eger \etal 2004).  Its formation time was significantly longer than Earth's, which may have consequences for its geological properties.

Systems with migrated giant planets tend to contain terrestrial planets both interior and exterior to their final orbits, although close-in terrestrial planets may not always survive (Fogg \& Nelson 2005).  "Hot Earth" planets tend to be formed interior to strong MMRs, especially the 2:1 MMR (Mandell \etal 2007).  These may be favorable for detection via transits (Croll \etal 2007) or transit timing (Agol \etal 2005; Holman \& Murray 2005).  Exterior terrestrial planets form on long timescales from scattered material and tend to be very water-rich for two reasons: 1) material scattered by the giant planet has very large eccentricities, causing large-scale radial mixing, and 2) gas drag causes more distant, icy planetesimals to spiral inward and delivery vast amounts of water to these planets (Raymond \etal 2006b; Mandell \etal 2007).  If the lifetime of the gaseous disk does not extend long past the end of migration (as in the scenario of Trilling \etal 1998), then exterior terrestrial planets could still form, but eccentricity damping would be slower (causing even longer formation times) and the amount of planetesimal in-spiralling would be reduced (somewhat reducing their water contents; Mandell \etal 2007).  Nonetheless, hot Jupiter systems may be good places to look for additional planets, including potentially habitable terrestrial planets.

\section{Terrestrial Planet Formation in the Known Exoplanet Systems}

Given the large number of simulations of terrestrial planet formation that have been done while varying the characteristics of giant planets, it is possible to apply certain limits to the known extra-solar planet population.  Specifically, we can constrain which of the systems of known (giant) planets may have formed currently-undetected terrestrial planets.  To accomplish this, I will follow the procedure employed in Raymond \etal (2006b) and described in detail in section 4 of Mandell \etal (2007), which divides giant planets into two classes: those interior and those exterior to the habitable zone.  I assume the habitable zone to be a simple function of the stellar luminosity, calibrated to a value of 0.8-1.5 AU for a Solar-mass star, in rough agreement with Kasting \etal (1993; for a more detailed treatment of the habitable zone, see Selsis \etal 2007b).  The stellar luminosity is calculated using the mass-luminosity relation of Scalo \etal (2007), which is a fit to the data of Hillenbrand \& White (2004).  Table 1 lists the inner and outer giant planet limits on the formation of a terrestrial planet in the habitable zone, reproduced from Mandell \etal (2007).

\begin{deluxetable}{ccccc} 
\tablewidth{0pt} 
\tablecaption{Giant Planet Limits for Potentially Habitable Systems} 
\tabletypesize{\scriptsize} 
\tablecolumns{5} 
\renewcommand{\arraystretch}{.6} 
\tablehead{
\colhead{M$_\star$ (M$_\odot$)} &  
\colhead{Sp. Type\tablenotemark{1}} & 
\colhead{Hab Zone (AU)\tablenotemark{2}} &  
\colhead{Inner Limit (AU)} & 
\colhead{Outer Limit (AU)}} 
\startdata
0.1 & M6 & 0.024-0.045 & 0.015 & 0.075\\
0.4 & M3 & 0.10-0.19 & 0.06 & 0.32\\
0.7 & K6 & 0.28-0.52 & 0.17 & 0.87\\
1.0 & G2 & 0.8-1.5 & 0.5 & 2.5 \\
1.3 & F8 & 2.3-4.3 & 1.45 & 7.2 \\
1.6 & F0 & 6.5-12.3 & 4.1 & 20.5\\
2.0 & A5 & 25-47 & 15.7 & 78.3\\
\enddata
 \tablenotetext{1}{Spectral types from Table 8.1 of Reid \& Hawley (2000) and Appendix E from Carroll \& Ostlie (1996). Note that spectral types of low-mass stars are age-dependent.}
\tablenotetext{2}{Habitable Zones calibrated to 0.8-1.5 AU for a solar-mass star.}
\label{tab:gplim}
\end{deluxetable}
   
For giant planets exterior to the habitable zone, I assume that terrestrial accretion occurred in a similar fashion to the Solar System, with the late stages of accretion governed in part by fully-formed giant planets.  In Raymond (2006), I ran 460 low-resolution simulations of terrestrial accretion under the influence of a single, Jupiter-mass giant planet.  Systems with giant planets beyond 2.5 AU formed terrestrial planets larger than 0.3 $\mearth$ within the habitable zone (0.8-1.5 AU), while only systems with giant planets beyond 3.5 AU formed $>0.3 \mearth$ water-rich terrestrial planets in the habitable zone.  These limits depend strongly on the eccentricity of the giant planet (as shown in Raymond 2006), and certainly also on the giant planet's mass, although this is not currently accounted for.  By assuming this separation (in terms of ratios of orbital periods) between the most distant terrestrial planet and the giant planet to be characteristic and independent of stellar mass, it was showed that only 5-6 \% of the known systems (as of 2006) could form Earth-like planets in the habitable zone in Solar System-like configurations (Raymond 2006).  Note that this prescription assumes that the giant planet acquired its large eccentricity before the bulk of terrestrial planet formation occurred.  If eccentricity was acquired later then the situation is more complicated: the giant planet's orbit may have been more favorable during terrestrial accretion, but the eccentricity-causing instability has the potential to destabilize the terrestrial planets (Veras \& Armitage 2006).

For giant planets interior to the habitable zone, limits are calibrated to simulations including giant planet migration (see $\S$ 4).  From the 8 relevant simulations of Mandell \etal (2007), the spacing between the close-in giant planet and the innermost large terrestrial planet varies considerably, from 3.3 to 43 in terms of orbital period ratios, with a median value of 9.  Thus, we adopt the median as a characteristic spacing, although there clearly exists a range of outcomes.  For the most optimistic case, a giant planet at 0.7 AU would allow a terrestrial planet to form just inside the outer edge of the habitable zone at 1.5 AU.  Note that we have limited calculations for inner giant planets on eccentric orbits.  We therefore require inner giant planets to have eccentricities less than 0.1, which was the largest eccentricity of any giant planet in the simulations of Mandell \etal (2007).  

\begin{figure}
\begin{center}
 \includegraphics[width=5in]{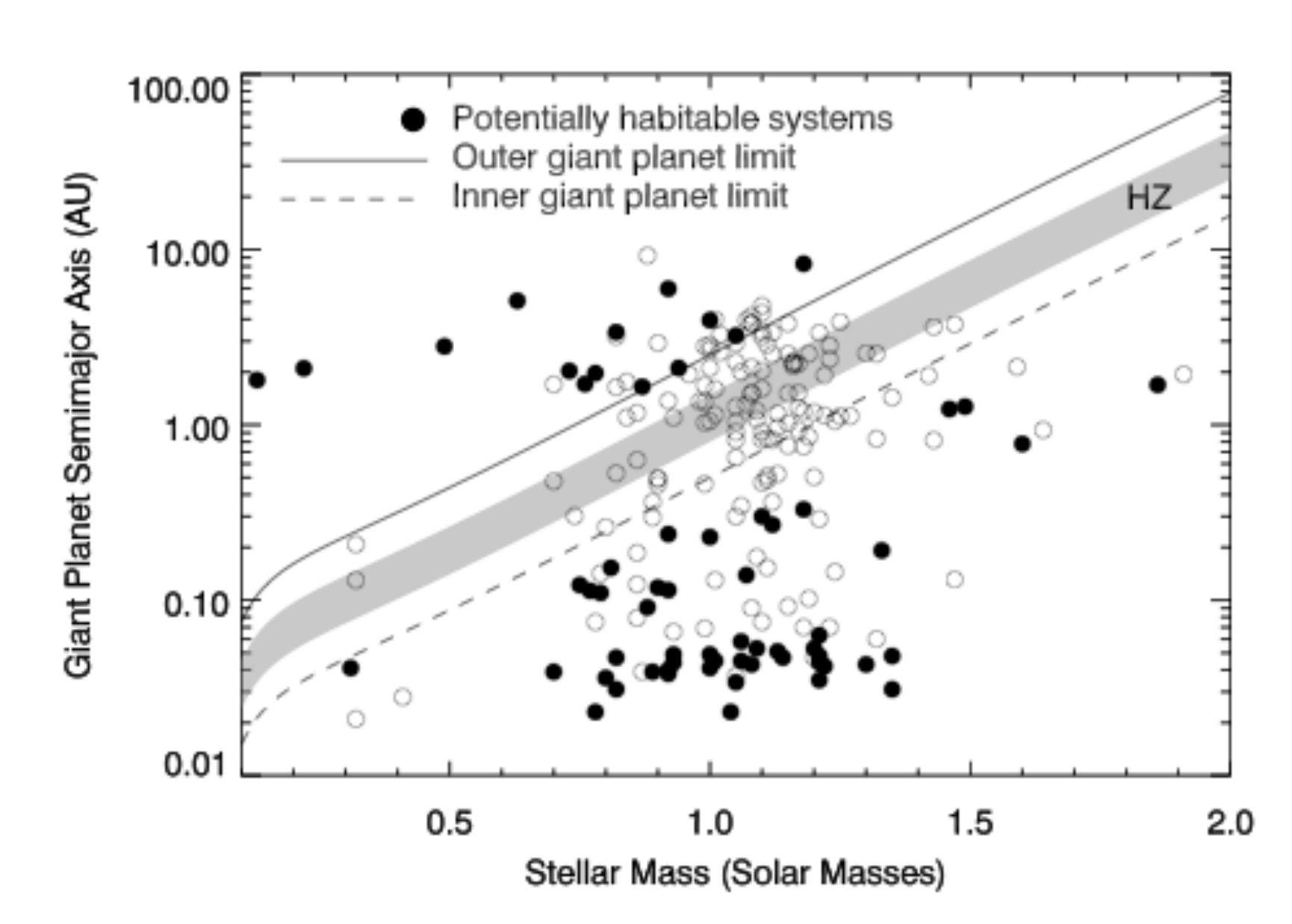} 
 \caption{Extra-solar planetary systems that could harbor an Earth-like planet in the habitable zone.  The habitable zone is shaded, and our inner and outer giant planet limits, listed in Table 1 for specific values of the stellar mass, are shown by the dashed (inner) and solid (outer) lines.  Each circle represents a known planet; those that fulfill either the inner or outer limit are filled in black.  Note that many planets that appear to meet the limits are not filled; this is because their orbital eccentricities were too large, although this is not shown explicitly.  From Mandell \etal (2007).}
   \label{fig:xsp}
\end{center}
\end{figure}

With knowledge of extra-solar planet host stars' masses, we can apply the limits from Table 1 to the know extra-solar planets.  I use the planet population as of August 1 2006, with values taken from Butler \etal (2006) and exoplanet.eu and references therein.  This sample includes 207 planets in 178 systems, with 21 multiple planet systems.  Figure~\ref{fig:xsp} shows the results of applying our limits to this sample of giant planets: 65 out of 178 systems (37\%) meet our criteria and appear able to have formed a terrestrial planet in the habitable zone.  Of these, 17 met our outer limits and could have Solar System-like architectures, while 49 could have giant planets interior to the terrestrial planets.  For a list of all systems that met our limits, see Mandell \etal (2007).  

One multiple planet system, 55 Cancri, met both our inner and outer giant planet limits.  Indeed, detailed simulations had already shown that an Earth-sized terrestrial planet could accrete {\it in situ}, in the presence of the four giant planets (Raymond \etal 2006c).  Interestingly, a fifth planet in 55 Cancri was recently discovered, although it is far larger than a terrestrial planet (Fischer \etal 2007).   In addition, our list was compiled prior to the discovery of the 5 $\mearth$ planet Gliese 581 c (Udry \etal 2007), whose orbit lies just interior to the habitable zone (although the more distant, 8 $\mearth$ planets Gliese 581 d may be within the outer boundary; Selsis \etal 2007b).  Indeed, based on the orbit of Gliese 581 b, the star was on our list of candidate systems for Earth-like planets. 

This is an interesting exercise in terms of its predictive power, an indication of which known systems are good candidates to search for Earth-like planets.  In addition, it is interesting to compare our results for where terrestrial planets can form with studies looking at where terrestrial planets' orbits would be stable.  Several authors have examined the stability of test particles, used as proxies for terrestrial planets, in the known systems (e.g., Menou \& Tabachnik 2003; Jones \etal 2005).  These studies find that about half of the known systems could have a stable terrestrial planet in the habitable zone, somewhat higher than our estimate that about 1/3 could form an Earth-like planet.  

It is important to realize that the existence of a stable region does not imply that a planet must occupy it.  For example, a planet of up to 5 $\mearth$ would be stable at 3 AU in the Solar System's asteroid belt (Lissauer \etal 2001), but it could not have formed there.  However, a contrasting example was recently discovered among the extra-solar planets, in the system HD 74156.  Two giant planets were known to exist in HD 74156 as of 2003 (Naef \etal 2004).  Detailed studies showed that both test particles and Saturn-mass planets were stable between the orbits of the two known planets, in a relatively narrow stable region (Barnes \& Raymond 2004; Raymond \& Barnes 2005).  However, simulations showed that accretion of smaller bodies could not occur in this stable zone, making it somewhat analogous to the Solar System's asteroid belt (Raymond \etal 2006c).  Recently, and with no prior knowledge of the theoretical results, Bean \etal (2007) discovered a $\sim$ Saturn-mass planet on the orbit predicted by Raymond \& Barnes (2005).  Thus, there exists an interesting contrast between the Solar System's asteroid belt (no planet accrete form there and there is no planet) with the previously-identified stable zone of HD 74156 (no planet could accrete there but there is a planet).  Perhaps this is simply due to the order of planet discoveries, and the stable zone where HD 74156 d is located has little meaning.  On the other hand, perhaps we have much more to learn about planet formation and dynamics.

\section{Conclusions}

In this proceedings, I have summarized the current state of knowledge about terrestrial planet formation in extra-solar planetary systems.  In $\S$ 2, I reviewed the stages of terrestrial planet formation, from micron-sized grains to planetesimals, protoplanets, and full-sized planets.  In $\S$ 3 I described the effect of external parameters such as the disk and giant planet properties on the terrestrial accretion process as well as radial mixing and water delivery.  In $\S$ 4, I presented recent results showing that terrestrial planets can form in systems with close-in giant planets, assuming those to have migrated to their final locations.  In fact, close-in giant planets should be accompanied by both very close-in "hot Earths" and exterior ocean planets (Raymond \etal 2006b). In $\S$ 5, I used results from previous work to derive limits designed to predict, based on the observed orbits of giant planets, where to search for other Earths among the known extra-solar systems. Approximately one third of the known systems of giant planets are good candidates for harboring an Earth-like planet (Mandell \etal 2007).

Many issues remain to be resolved in each topic I described.  The details of planetesimal formation are still poorly known, although recent results suggest that several process including turbulence and migration of meter-sized bodies acting in tandem might be the solution (Johansen \etal 2007).  The details and consequences of giant impacts are not well understood, in terms of the fate of collisional debris and compositional changes induced by the impacts (Genda \& Abe 2005; Asphaug \etal 2006; Canup \& Pierazzo 2006).   In addition, some of the effects of external parameters are perhaps less well understood than we would like to think.  For example the giant planet eccentricity - terrestrial planet water content correlation (Chambers \& Cassen 2002; Raymond \etal 2004) inherently assumes that eccentric giant planets acquire their eccentricities early, before terrestrial accretion.  If this assumption is wrong, then one can imagine a scenario in which  terrestrial planets tend to form with relatively circular giant planets; this is a beneficial situation in terms of water delivery.  However, a late, impulsive eccentricity increase can destabilize the orbits of terrestrial planets and remove them from the system entirely (Veras \& Armitage 2006).  Indeed, there are many known extra-solar planets that would be favorable for terrestrial planet formation if their orbits were more circular (see Fig.~\ref{fig:xsp}).  Thus, if giant planet eccentricity is acquired late, many systems may undergo ``planetary system suicide'', forming an Earth-like planet and subsequently destroying it.  A more complete, holistic view of planet formation and evolution is needed to distentangle these effects.

\section{Acknowledgments}

I want to thank the organizers of the symposium for an excellent meeting in a beautiful place.  The research presented here was made possible by contributions from several of my collaborators, including Tom Quinn, Jonathan Lunine, Avi Mandell, Rory Barnes, Steinn Sigurdsson, John Scalo, Vikki Meadows, Nate Kaib, Eric Gaidos, and Nader Haghighipour.  I was supported by an appointment to the NASA Postdoctoral Program at the University of Colorado Astrobiology Center, administered by Oak Ridge Associated Universities through a contract with NASA.  I dedicate this paper to my son Owen Zahler Raymond, born January 1, 2008.

\end{document}